# IFTT-PIN: A PIN-Entry Method Leveraging the Self-Calibration Paradigm

We present 'IF This Then PIN', a PIN-entry method whose code-pad configuration only pre-exists in the user's mind and is learned on-the-fly by the interface.

Dr. Jonathan Grizou

School of computing, University of Glasgow

Center for Research and Interdisciplinarity (CRI). Université de Paris

IFTT-PIN is a self-calibrating version of the PIN-entry method introduced in Roth et al. (2004) [1]. In [1], digits are split into two sets and assigned a color respectively. To communicate their digit, users press the button with the same color that is assigned to their digit, which can thus be identified by elimination after a few iterations. IFTT-PIN uses the same principle but does not pre-assign colors to each button. Instead, users are free to choose which button to use for each color. The button-to-color mapping only exists in the user's mind and is never directly communicated to the interface. In other words, IFTT-PIN infers both the user's PIN and their preferred button-to-color mapping at the same time, a process called self-calibration. In this paper, we present online interactive demonstrations of IFTT-PIN with and without self-calibration and introduce the key concepts and assumptions making self-calibration possible. IFTT-PIN can be tested at https://jgrizou.github.io/IFTT-PIN/ with a video introduction available at https://youtu.be/5I1ibPJdLHM. We review related work in the field of brain-computer interface and further propose self-calibration as a novel approach to protect users against shoulder surfing attacks. Finally, we introduce a vault cracking challenge as a test of usability and security that was informally tested at our institute. With IFTT-PIN, we wish to demonstrate a new interactive experience where users can decide actively and on-the-fly how to use an interface. The self-calibration paradigm might lead to novel opportunities for interaction in other applications or domains. We hope this work will inspire the community to invent them.

## 1 INTRODUCTION

Interfaces help users convey their intent to machines. A pre-requisite of successful interaction is for both the user and the machine to agree on how the interface through which they communicate works. Specifically, actions performed by the user on the interface (e.g., pressing this or that button) should be interpreted the same way by both the user and the machine. This action-to-meaning mapping is usually pre-defined by design (each button has a well-defined function) and learnt by users through various onboarding experiences. For example, a standard remote control is a pre-defined interface. Each button is clearly labelled with their function and an instruction manual is available to teach users if required.

In some cases, the action-to-meaning mapping cannot be pre-defined or would gain from being configured by the user. For example, in brain-computer interfaces (BCI), the “brain-signal-to-meaning” mapping cannot be pre-defined because each user has a different “brain signature” which needs to be learnt before any meaningful interaction can take place.

In practice, such personalization requires an explicit calibration procedure to learn the action-to-meaning mapping of each user. During calibration, the user is tasked to produce examples of action-meaning pairs, which are used as a reference to learn a tailored action-to-meaning mapping. For example, in brain-computer interfaces, users are asked to think about moving their left or right hand while their brain signals are being recorded. Based on these examples, using machine learning tools, it is possible to build a “brain-signal-to-meaning” model for this particular user.

Such calibration procedure is required due to the chicken-and-egg nature of the problem. To understand what a user is wanting to do, we need to be able to interpret their actions. But to interpret their actions, we first need to know what the user is trying to do in order to build an action-to-meaning model.

A curious person might ask: “Could we circumvent this chicken-and-egg problem? Could we allow users to start controlling a machine using their preferred action-to-meaning mapping and without knowing it in advance?” An interface capable of solving this problem can be defined as self-calibrating. Interestingly, self-calibrating interfaces have been studied before, mainly in the BCI community where the chicken-and-egg problem is made salient due to the inter-user variability of brain signals. We will review these works in section 7 and rather focus here on the interactive opportunities unlocked by self-calibrating interfaces.

A self-calibrating interface implies that the action-to-meaning mapping initially only exists in the user’s mind. In other words, actions performed by the user only make sense to them. An outside observer would not be able to interpret the user’s actions. Watching someone use a self-calibrating interface would be like watching someone using a remote control with no labels on the buttons. This alone represents a defense mechanism against shoulder surfing attacks where malicious observers try to infer user personal information or password by watching other’s actions 'over their shoulder’.

In this paper, we present an application of self-calibrating interfaces to a PIN-entry task and show its potential use as a defense mechanism against shoulder surfing attacks. We based our demonstrator on the cognitive trap door

PIN-entry method presented by Roth et al. in [1], which makes use of two colored buttons to selectively refine the user password via an elimination process. In [1], those buttons are clearly labelled by color and an outside observer could easily infer the sequence of colors entered by the user. By introducing self-calibration, we can remove the colors on the buttons and allow the user to choose which color to attribute to each button in their mind, thus making it much harder for an outside observer to infer the user's PIN.

We call our method IFTT-PIN, which stands for 'If This Then PIN', because it makes extensive use of simple 'If This Then That' reasoning. When buttons are of known colors, the logic goes as follow: *"If the user pressed the button B, then they indicated that their digit is of color C, thus their digit is among the set of digits currently colored in C."* The self-calibrating version flips this reasoning on its head and on a per digit case: *"If the user is trying to type the digit D, which is currently colored in C, then when they used the button B, they meant the color C"*, which, when combined with a test of user consistency: *"If, for a particular digit D, the user pressed the same button B to mean two different colors (C1 and C2), then they are not entering the digit D"*, enables us to solve the self-calibration problem. This reasoning will be explained in detail with illustration in section 4. All our interactive demos are available online with links in section 3.5 and 4.7, including additional tutorial versions that shows the algorithmic and decision-making process in a side panel.

In what follows, we first present our PIN entry interface with buttons of known colors. We then explain the main concepts behind self-calibration and how it is implemented in IFTT-PIN. Once the reader is familiar with IFTT-PIN, we discuss how it could be used as a defense mechanism against shoulder surfing attacks and present a vault cracking challenge as a usability and security test that we tested at our institute. Only then, in section 7, we discuss related work in Human-Computer Interaction (shoulder surfing, motion matching, adaptive interfaces), Brain-Computer Interfaces (P300 speller, self-calibration), Human-Robot Interaction (interactive learning), and language acquisition (experimental semiotics and language games). Finally, we identify several limitations and highlight promising research directions for self-calibrating interfaces.

Beyond the specific IFTT-PIN implementation, our hope is to trigger curious minds in exploring other applications or domains that could benefit from self-calibration that might lead to novel interaction paradigms.

## 2 CONTRIBUTIONS

- A new PIN entry experience leveraging self-calibration to allow users to choose the meaning of each button on-the-fly without calibration.
- A novel approach to overwhelm the observer in shoulder surfing scenarios relying on on-the-fly code-pad configuration.
- The description of a vault challenge as an informal usability and security test for shoulder surfing defensibility.
- The first demonstration of a self-calibrating interface using discrete user actions (i.e., using buttons)
- An online interactive demonstrator that allows to explain the self-calibrating paradigm in a few minutes anywhere on a computer or smartphone.

## 3 PIN-ENTRY METHOD

While designed independently, IFTT-PIN follows the same principle as the PIN-entry method from Roth et al. [1]. Quoting directly from [1]: *"The principal idea is to present the user the PIN digits as two distinct sets e.g., by randomly coloring half of the keys black and the other half white. The user must enter in which set the digit is by pressing either a separate black or white key. Multiple rounds of this game are played to enter a single digit and it is repeatedly played until all digits are entered. The verifier e.g., the automatic teller machine (ATM), determines the entered PIN digits by intersecting the chosen sets."*.

IFTT-PIN differs as follows: (1) we use grey/yellow instead of black/white, (2) digits are organised differently, in two rows instead of as a phone pad, (3) our interface actively chooses which color to apply to each digit at each turn instead of following a pre-existing sequence, and (4), in section 4, we implement self-calibration that removes the need to pre-define the color of each feedback button. This section is meant to familiarize the reader with our interface without the self-calibration element and introduce some basic vocabulary. The next section presents our contribution.

### 3.1 Interface design

As shown in Figure 1, we split our interface into three parts. The top part displays the PIN. The middle part shows all possible digits (from 0 to 9) colored in yellow or grey according to which set they belong to. Think of this section as the machine asking the user: "What color is the digit you want to type?". The bottom part is dedicated for the user to answer that question. In the case of [1], the user feedback takes place via two colored buttons. Here the left button is yellow, and the right button is grey. Think of this section as the human answering to the machine: "My digit is yellow" or "My digit is grey".

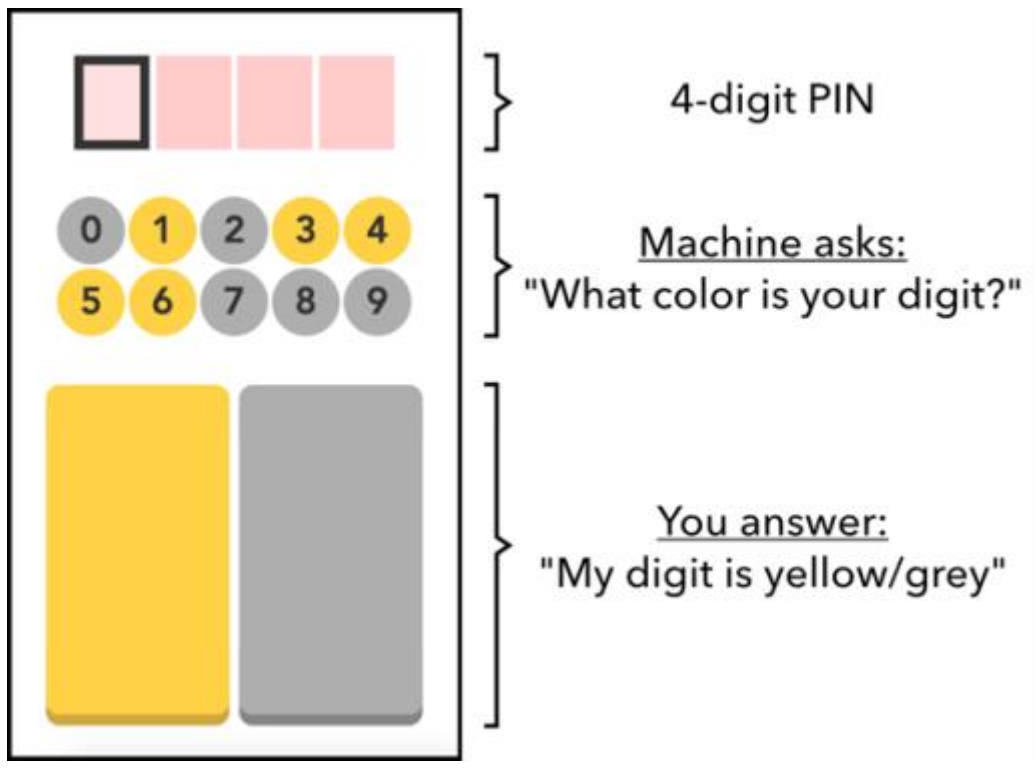

*Figure 1: Breakdown of the IFTT-PIN interface*

### 3.2 Vocabulary

As illustrated in Figure 2, we will refer to the following elements. An **action** is what the user does in order to convey their meaning, here, the user's actions are to press either the left or the right button. A **meaning** is what the user wants to say to the machine, here it is either: "My digit is yellow" or "My digit is grey". An **intent** is what the user wants the machine to do, here entering a specific PIN, one digit at a time. To put it simply, an action conveys a meaning that is used to infer an intent. But inferring an intent from a meaning requires a bit of context about the task, protocols, and overall aim of the interaction.

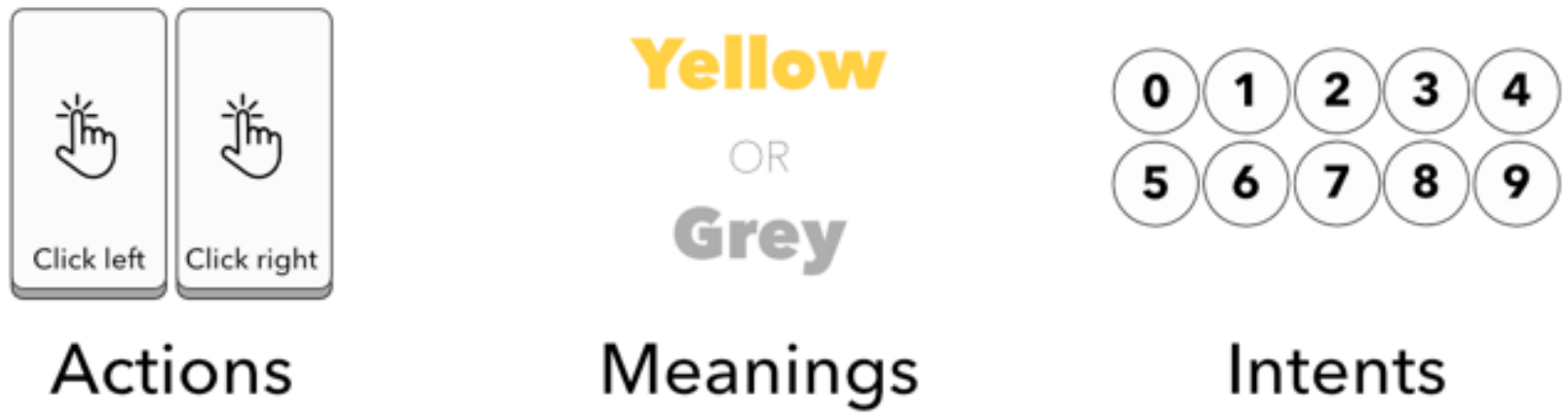

*Figure 2: Elements of language used in this work.*

### 3.3 Assumptions

The context for IFTT-PIN is a user wanting to type a PIN via this specific interface. More specifically, we assume that: (1) the user aims to type a PIN one digit at a time. Thus, their current intent is to type one of ten possible digits. (2) The user follows the established convention of indicating the color of the digit they want to type. Thus, the possible meanings are yellow or grey. (3) The user can perform one of two actions, pressing either the left or the right button. This is constrained by the design of our interface. (4) The mapping between the user's actions and their meanings is known. Pressing the left button conveys the meaning yellow and the right button conveys the meaning grey, see Figure 3.

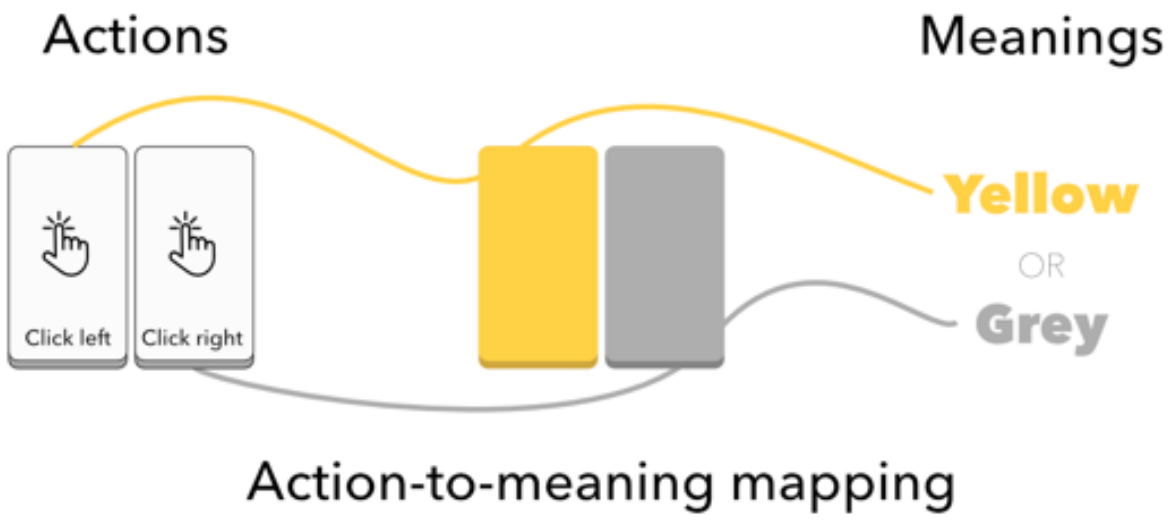

*Figure 3: Illustration of the user action-to-meaning mapping used in this section.*

The prior knowledge of the action-to-meaning mapping is the most relevant assumption here because it will be removed in the next section. For now, a known action-to-meaning mapping means that there is a pre-existing shared understanding between the user and the machine about the meaning conveyed by the pressing of each button. This information is made salient to the user by the colors displayed on each button, the left button is yellow, and the right button is grey.

### 3.4 Algorithmic principle

Knowing this mapping, and given the context of the PIN entering task, the interface identifies the digit the user wants to enter by reasoning as follows: *"If a user presses the left button (action), then it indicates that their digit is currently yellow (meaning), thus their digit is among the yellow-colored digits and all the grey digits can be discarded (intent).*", see Figure 4.

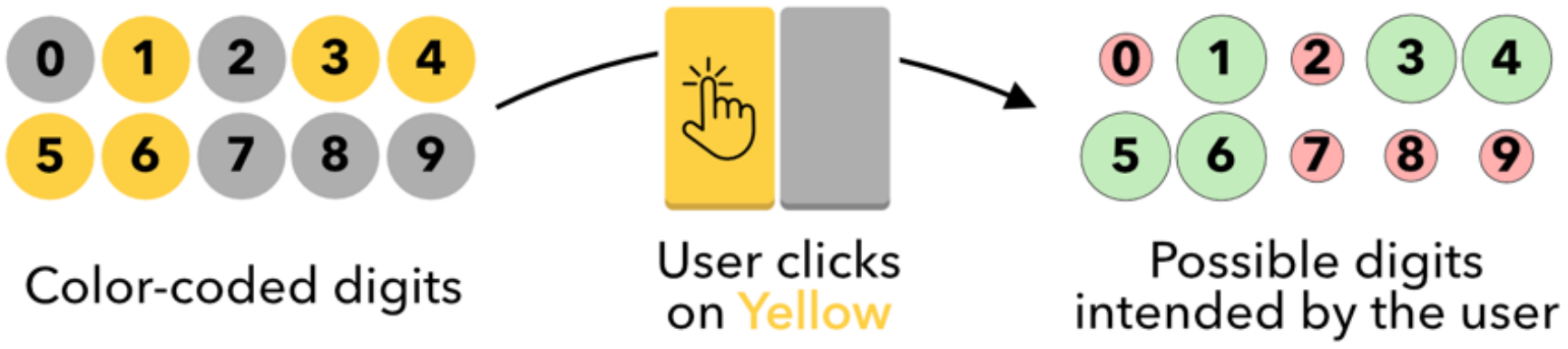

*Figure 4: One step of the inference process when the action-to-meaning mapping is known.*

By iteratively changing the color applied to each digit, we can narrow the possible digits down to the one the user has in mind. This is the same process as described in Roth et al. and an interested reader can refer to their pseudocode for a more formal algorithmic description using sets (see Figure 2 in [1]).

### 3.5 Interactive demonstration

The interface described above can be used online at https://jgrizou.github.io/IFTT-PIN/interaction_1.html. To visualize the decision process while using the interface, we developed a version with a side dashboard that displays the history of your clicks with respect to each digit. Each click is shown as a dot on the button clicked by the user and colored of the same color than the associated digit was when the user clicked on it. If the color of the dot differs from the color of the button, we can discard the associated digit as not being the one the user intends to enter. Because there are 10 possible digits, we are showing 10 individual panels, one for each digit. This version is available at https://jgrizou.github.io/IFTT-PIN/interaction_1_sidepanel.html. If preferred, a walkthrough video demonstration can be found at https://youtu.be/6wgOa380uEo.

Having understood and played with this PIN-entry method, we remind the reader that our contribution lies in the introduction of self-calibration to this interface, which we describe below.

## 4 SELF-CALIBRATION METHOD

We now remove the assumption that an action-to-meaning mapping is known. This means that buttons have no predefined colors. Instead, the user decides the colors of each button in their mind and uses them as such - never explicitly telling the machine about it. The machine will nonetheless be able to identify the user PIN along with the colors of the buttons.

First, we need to understand that the reasoning used in section 3.4 does not work anymore. More specifically, the following link in the chain of inference is broken: "If a user presses the left button (action), then it indicates that their digit is currently yellow (meaning).", because we do not know the color attached to each button. The usual approach for solving this problem is to calibrate the interface, that is to learn the action-to-meaning mapping in order to repair the inference process.

This can be done in two ways by directly asking the user to assign colors to each button using another interface, for example using a digital paint brush, or by indirectly asking the user to perform a known task which will inform us of the color they assign to each button. For example, by enforcing which digit is to be entered, say a 1, we can reverse the inference pipeline and use the following reasoning: "Knowing that the user is typing a 1, if the digit 1 is yellow and the user is pressing the left button, then the left button means yellow." - hence learning the action-to-meaning mapping. This approach highlights the chicken-and-egg problem we are facing. To infer the user's digit, we need to know the button-to-color mapping. But to infer the button-to-color mapping, we need to know the user's digit.

In this work, we do not rely on any of the above solutions and resolve the chicken-and-egg problem by leveraging the self-calibrating paradigm developed in the BCI community (see related work in section 7.4.2). Specifically, we expose and exploit an assumption hidden in most interactive systems.

### 4.1 Consistency assumption

When using the interface with colored buttons from section 3, we automatically assume that a button has one and only one meaning, and that this meaning is the color displayed on each button, either yellow or grey (never both). This assumption is hard to formulate when colors are visibly assigned to each button, it is almost too obvious to be noticed. Generalizing from this particular "yellow or grey" case, we can formulate a last assumption in our PIN-entry task: (5) the user is assumed to be consistent in its usage of the interface, where consistency is defined as one button can only express one color.

### 4.2 Self-calibration principle

How can this new assumption help us? Because consistency is something we can observe. For example, if we ask a user to type a specific digit, say 1, and the user presses the same button twice both when the digit 1 is yellow and again when it is grey, then we can confidently say that the user is being inconsistent with its use of the interface.

But when facing the self-calibration problem, we do not know which digit the user is entering. However, the context given by the PIN entering task is well defined and includes a strong constraint: we know that the user is entering one of the ten digits from 0 to 9. We can thus formulate ten different hypotheses, according to which we can interpret the user's action. We call them interpretation hypotheses.

We also know that only one hypothesis is valid, the user is trying to type only one digit. In other words, the user can only be consistent with one hypothesis and will, in the long run, invariably breach consistency when evaluated according to the other hypothesis. By measuring the consistency of the user according to each interpretation hypotheses, we can discard the ones that show breach of consistency. This combined use of a consistency assumption and interpretation hypothesis is the basis of self-calibrating systems.

### 4.3 Algorithmic principle

Measuring consistency requires us to keep track of past interactions and to interpret them in light of each possible digit ($d \in D$) the user might be intending to enter. Specifically, we keep the history of users button presses ($b \in B$) and digit colors ($c \in C$) and run the following reasoning for each digit: *"If the user is trying to type the digit $d$, then when they used button $b$, they meant color $c$."* For each digit $d$, the algorithm is building a history of $(b, c)$ pairs representing samples from the action-to-meaning mapping if the user was entering $d$. For a given digit $d$, this historical sequence can be written as $H_N^d = \{ (b_i, \ c_i \mid d), \ i = 1, \dots, N \}$ where $b_i$ and $c_i \mid d$ represent respectively the button pressed by the user and the color of the digit $d$ at iteration $i$ out of N iteration performed so far.

Under our consistency assumption, a user cannot use the same button to express more than one color. For each digit and button, this reasoning unfolds as: *"If a user is trying to enter digit $d$, and used the button $b$ to mean both $c^{yellow}$ and $c^{grey}$, then the user would have been inconsistent and is thus not trying to enter d."* This is the core reasoning behind IFTT-PIN.

Assuming no human error, Boolean logic amply suffices to test for consistency. Specifically, if the number of colors assigned to a button is greater that one, then the user is being inconsistent. We can define $H_N^{d,b} = \{\ (\ c_i \mid d)\ if\ b = b_i,\ \ i =\ 1, \dots, N\ \}$ a subset of $H_N^d$ containing only the history of colors from $H_N^d$ that are linked with the button $b$. If, for a given digit $d$, for all $b$, the cardinality (number of unique elements in a set, here number of unique colors) of $H_N^{d,b}$is greater than 1, then the user is inconsistent. Using mathematical notation, for a given $d$, inconsistency is defined as $\exists b \in B, |H_N^{d,b}| > 1$, and, conversely, consistency can be written as $\forall b \in B, |H_N^{d,b}| \leq 1$. In turn, the consistency assumption could be expressed using the uniqueness quantification notation as $\exists! d \in D, (\forall b \in B, |H_N^{d,b}| \leq 1)$, which can be read as: "there is exactly one digit for which, for all buttons, the number of colors assigned to each button was at most one ".

This uniqueness is satisfied only once the user has "interacted enough" with the interface. It is valid at the limit when $N \to \infty$ and enough "variety of color pattern" has been displayed on the digit. In other words, only one unique hypothesis will eventually remain consistent. In practice, in most cases, the user digit can be identified after only a few clicks as the reader will be able to test via the interactive demo linked in section 4.7.

In summary, to identify the user digit, IFTT-PIN measures consistency for all digits after each iteration. Once only one digit satisfies the consistency requirement, we know it is the one the user has in mind. A pseudocode of this process is summarized in Figure 5 using a Pythonic syntax.

```
1   nDigit = 10; nButton = 9
2   # init history of interaction H_N^{d,b}
3   historyPerDigit = [ []*nButton ]*nDigit # empty at start
4   # init list of possible digits the user has in mind
5   consistentDigits = list(range(nDigit)) # all digits are possible at start
6
7   # repeat until only one user digit is consistent
8   while (len(consistentDigits) > 1):
9       digitColors = applyColorToDigits() # return list of color for each digit
10      buttonPressed = waitForUserInput() # return button pressed by the user
11      # store history of color per button per digit
12      for digit in range(nDigit):
13          historyPerDigit[digit][buttonPressed].append(digitColors[digit])
14      # test for consistency for each digit
15      consistentDigits = [] # empty list of consistent digit
16      for digit in range(nDigit):
17          # tmp variable, assumed consistent unless prooven otherwise
18          isDigitConsistent = True
19          # check consistency button per button
20          for button in range(nButton):
21              # compute the set of unique color attached to this button
22              colorSet = set(historyPerDigit[digit][button])
23              # inconsistent if more than one color in the set
24              if (len(colorSet) > 1):
25                  isDigitConsistent = False
26          # if digit consistent, we add it to the consistent list
27          if (isDigitConsistent):
28              consistentDigits.append(digit)
29
30  # consistentDigits[0] is the digit the user has in mind
31  # historyPerDigit[consistentDigits[0]] contains button-to-color mapping info
```

*Figure 5: Pseudocode for IFTT-PIN. The history of button presses and digit colors are stored in a variable called historyPerDigit (L12-13). A test of consistency is performed (L14-25) and all consistent digits are added to the*

*consistentDigitis list (L27-28). The process stops when only one consistent digit remains in the list (L8), which is the digit the user has in mind (see L30-31). The implementation details for applyColorToDigits() (L9) and waitForUserInput() (L10) may vary per application and are not core to the self-calibration method (although some variability in the digit colors is expected, see section 4.6).*

### 4.4 Algorithmic illustration

To adequately demonstrate IFTT-PIN, we decided to increase the number of buttons from 2 to 9 (Figure 6). This increases the possible color patterns from 2 to 510 which helps when demonstrating the interface.

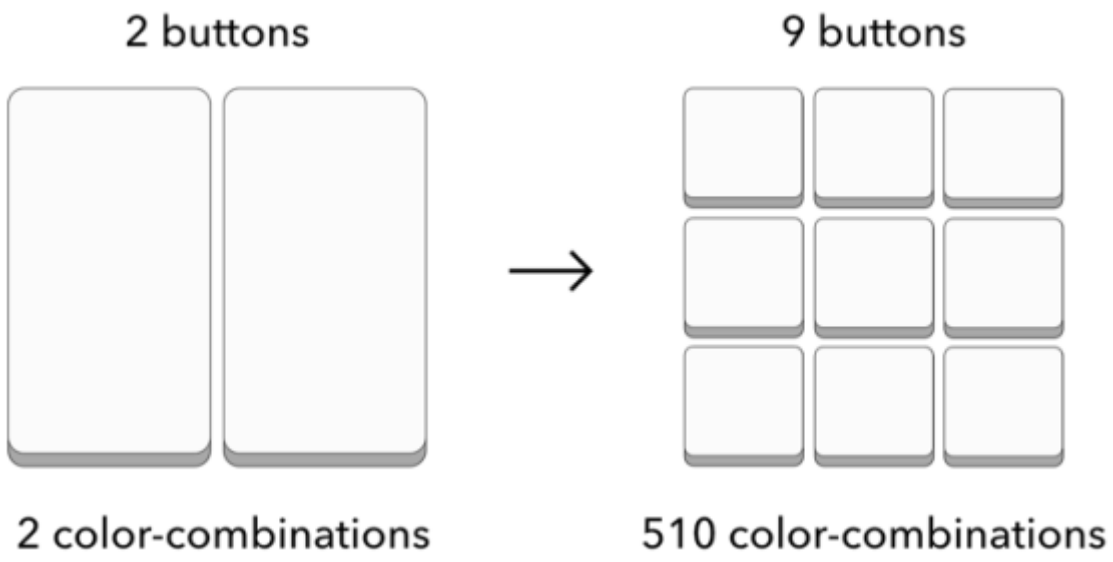

*Figure 6: For the self-calibration version of IFTT-PIN, we increased the number of buttons from 2 to 9. This allows us to define more varied color patterns when challenging the machine.*

Because the color of each button is now chosen by the user in their mind, buttons are shown in black on the interface to indicate that they do not have a color attached to them yet. As a user, you simply decide a color pattern to apply on the buttons and use the buttons that way. For example, the top three buttons could be yellow and all the others grey. See Figure 7 for more examples.

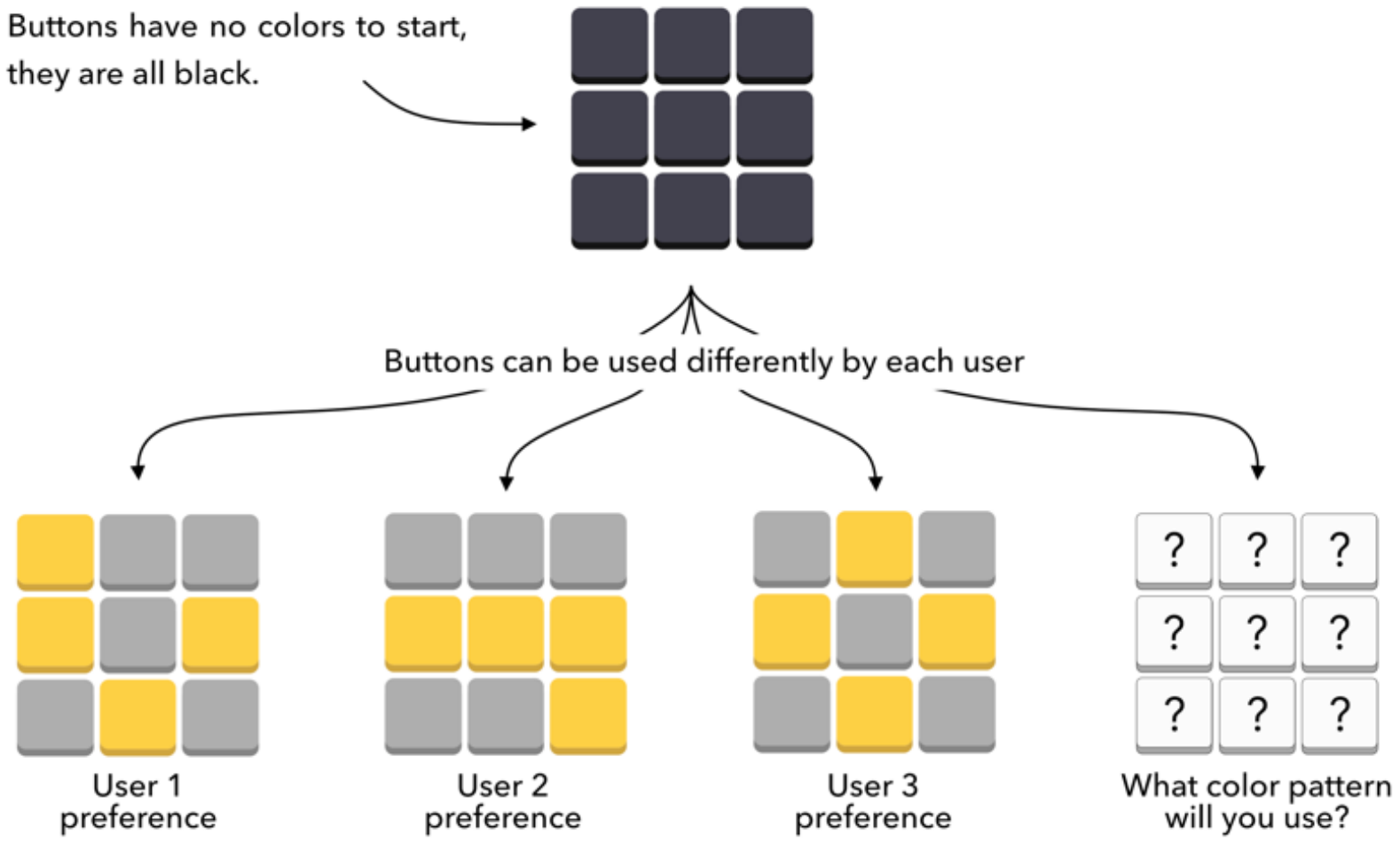

*Figure 7: Example of choice of button-to-color mapping. Each user can assign colors to each button as they prefer and use them as such. At least one button should be assigned for the yellow and grey colors.*

With this in mind, the interpretation hypotheses process unfolds as follows. First, colors are displayed on each digit. Then, you look at the digit you want to enter and click a button of the same color (following the button-to-color mapping you have defined in your mind). Each time you click on a button, the algorithm records which button you have clicked and what color was assigned to each digit at the time you clicked. The algorithm is then making an interpretation hypothesis for each digit that we can visualize by adding a small dot on each button used and assigning it the same color as the digit was when you clicked on the button.

This process is illustrated in Figure 8 after one, four, and eight clicks from a typical interaction. For visual clarity, we only show the process for digit 0, 1, 2, and 3, as if the user could only enter one of those four digits. After one click, the top-left button is marked with a dot because the user just clicked on that button. This dot is yellow for digit 0 and 3, and grey for digit 1 and 2, because, at the time the user clicked on the button, the digits 0 and 3 were yellow, and digits 1 and 2 were grey. After one click, there is no way to know which digit the user is entering, nor what color is assigned to that button, all hypotheses are consistent. After four clicks, the middle button has received two clicks from the user. If the user was entering digit 0 and 2, they would have used the middle button to mean alternatively yellow and grey, which would be a breach of the consistency assumption. Thus, the user is not entering digit 0 or 2. For digit 1 and 3, both times the user clicked on the middle button to mean the same color. The middle button would be yellow if the user was typing a 1, and grey if the user was typing a 3, but both options remain consistent at this stage, and the user might well be entering any of them, we just can't tell yet. After eight clicks, the top-left button would have been used for both yellow and grey if the user was entering digit 1. Only our interpretation of the button color according to digit 3 remains fully consistent, thus we can conclude that the user is trying to enter the digit 3. This process is best understood interactively, and the same visual representation from Figure 8 is available as an interactive demonstration (see link in section 4.7).

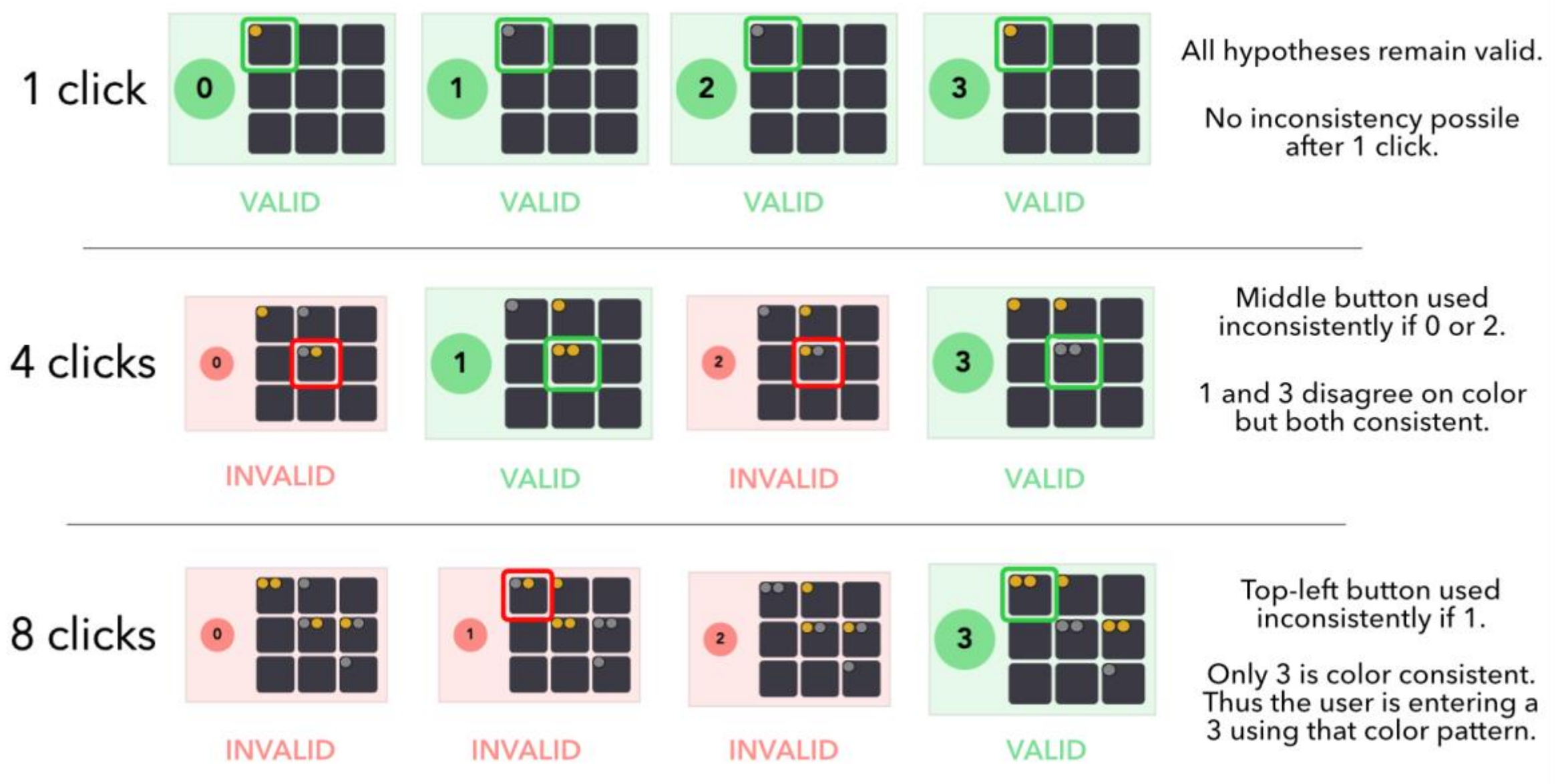

*Figure 8: Illustration of inconsistency detection for digits 0 to 3 after one, four, and eight clicks from a typical interaction. After each iteration, a dot is placed on the button pressed by the user and is colored of the same color as was the color of*

*the digit when the button was pressed. Green squares higlight buttons of interest for which hypothesis is consistent. Red ones highlight inconsistencies, meaning the same button would have been used to mean two different colors. Notice how none of the hypotheses share the same button-to-color mapping, yet several mappings can remain consistent for many steps. For example, after 4 clicks, hypothesis 1 and 3 disagree on the color to assign to the middle button. Yet, in both cases, the usage of the button is consistent and thus both hypotheses remain valid.*

### 4.5 Learning the action-to-meaning mapping

It is important to understand that once we have identified a digit, we automatically know which color the user is attributing to each button. In other words, the process of entering the first digit under the self-calibration paradigm is calibrating the system for subsequent use by the same user. Within the IFTT-PIN interface, we show this by assigning colors to each button once the interface has identified a new digit. When the user reuses the same buttons to enter a second digit, it will therefore take less iteration than when entering the first digit by falling back to the reasoning from section 3.4. For buttons that remain undefined (black), we keep using the self-calibration algorithm.

### 4.6 Active choice of digit color

The choice of the color that are applied to each digit at each interaction step is important to narrow down to the correct hypothesis. In [1], the sequence of color was pre-defined and ensured that the digit could be uniquely identified. In IFTT-PIN, we are actively deciding which color to assign to each digit at each iteration according to the history of interaction and in order to identify the user intent as fast as possible. For self-calibration problems, an active planning strategy has been proposed in [2] which we implemented in IFTT-PIN. We added an additional constraint that each colored set of digits must be balanced at each iteration, in other words, there must always be five yellow digits and five grey digits. The implementation details are beyond the scope of this paper, but our code is available online for inspection at https://github.com/jgrizou/openvault.

### 4.7 Interactive demonstration

The self-calibrating IFTT-PIN interface can be tested online at https://jgrizou.github.io/IFTT-PIN/interaction_2.html. As before, we developed a version with a side dashboard that displays the history of clicks with respect to each digit, which is available at https://jgrizou.github.io/IFTT-PIN/interaction_2.html. If preferred, a walkthrough video demonstration can be found at https://youtu.be/t7MQoBnzryQ.

## 5 LIVE DEMOS AND MAGIC TRICKS

IFTT-PIN is particularly effective for live demos. First, all interactive demos are freely available online and do not require any installation, they work on both computer and smartphone and can thus be demonstrated anywhere in under one minute. Additionally, PIN entering is a familiar task, performed daily by most people, which makes it easy to give context for the demonstration. For guaranteed effect, we highly recommend mimicking a magician's act by asking different members of the audience to publicly choose a PIN and the colors to assign to each button. You can then enter the chosen PIN using the chosen color pattern in front of the audience. After a few clicks, both their PIN and their colors will appear on screen, as per magic, leaving the audience mesmerized. Having grasped their attention, and maybe after a few additional demonstrations with other PIN and color combinations, you can proceed to explain how the interface works using the tutorial version with the side explanatory panel.

## 6 USABILITY AND SAFETY

To test both the usability and safety of IFTT-PIN, we initiated a vault opening challenge within our institute. We built a transparent vault that can only be opened by entering the correct PIN via IFTT-PIN. The vault was installed into the common area of our institute and treats were placed into the vault visible to all. To test the robustness to shoulder surfing attacks, we released a video of someone opening the vault with the PIN hidden. Students and staffs were then free to attempt to crack the code, open the vault, and collect its content. As shown in Figure 9, this triggered interest with a large number of students who actively engaged with the game.

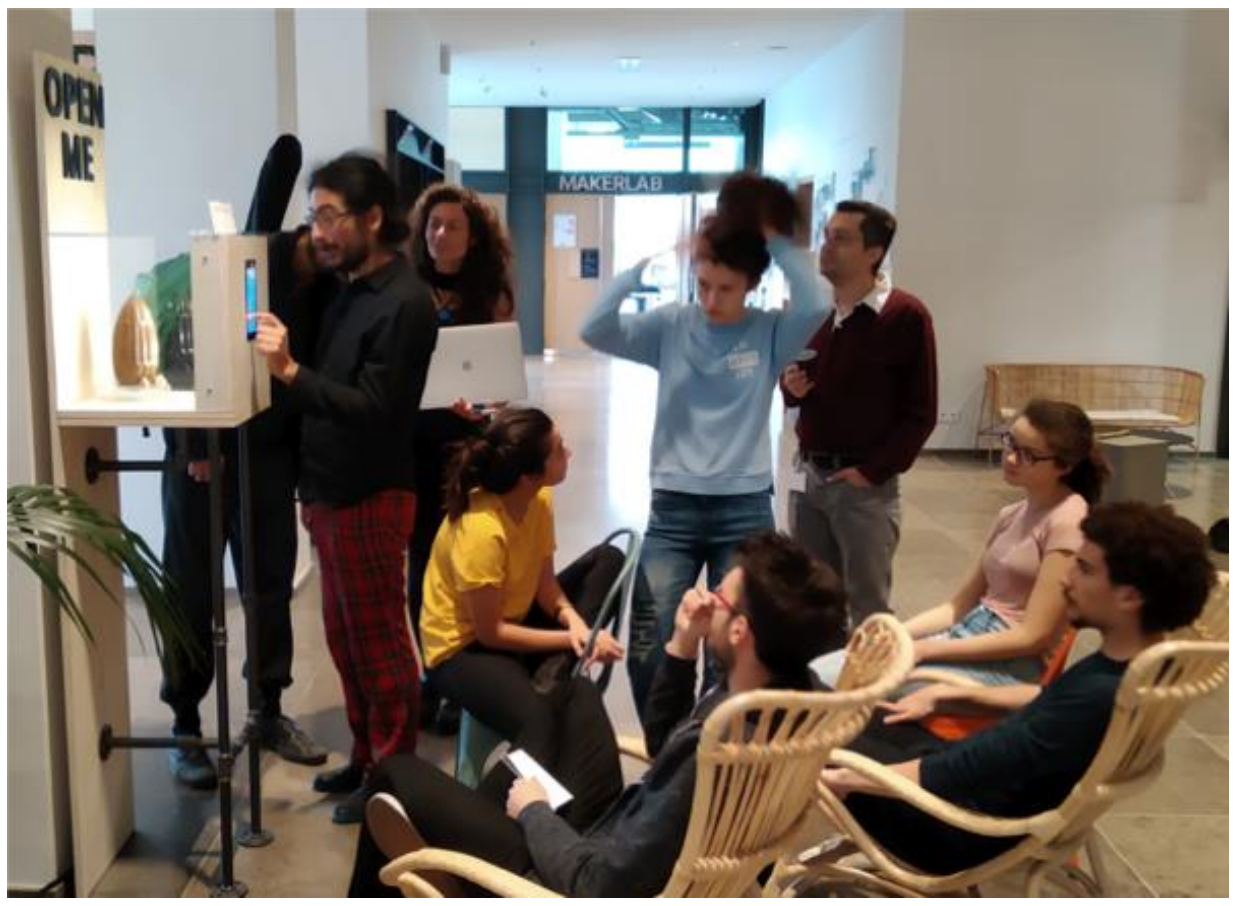

*Figure 9: Vault opening challenge at our institute. A chocolate egg is placed in a transparent vault protected by IFTT-PIN. A video of the author entering the correct PIN was made available online. Students and staffs needed to crack the PIN and enter it correctly to earn the right to eat the chocolate egg.*

To make the challenge more gradual, we started with the two-button interface from section 3 and progressively increased complexity to the full self-calibrating version of the interface that did not show the code, nor the button's color, even after the PIN was entered by the user. Having read this paper, the interested readers can challenge themselves on that final level by watching the video at https://youtu.be/QNDrZ7tvdb8 and verifying if they cracked the code at https://jgrizou.github.io/IFTT-PIN/challenge.html.

Over several days, as difficulty increased, students were voluntarily forming groups to crack the code. This was entirely on their free time and not part of any official curriculum activities. After hours of work, students got the hang of the problem and treats were slowly disappearing from the vault.

This give us three important pieces of information: (1) IFTT-PIN provides a good defense against shoulder surfing attacks as it required time, effort and access to a video recording to be cracked. The fact that IFTT-PIN is not fully shoulder-surfing-proof is not surprising because the self-calibrating algorithm works using only the information shown on screen, which is the same information accessible to an outside observer. No secret communication channel exists with the user, thus an observer with time and full visibility can run the same procedure and infer the PIN. This decoding is however extremely challenging to do in real time for a human due to cognitive overload, (2) IFTT-PIN is usable because participants were able to enter the correct PIN and open the vault to collect its content without access to a tutorial or explicit instruction, and (3) the vault challenge is engaging, challenging and self-motivating, thus showing promise as an educational and experimental setup. By cracking the code, people of various ages (from 8-year-olds to adults) learned various computer science concepts and reinvented for themselves the consistency principles used to solve the self-calibration problem that was only recently introduced in academic circles.

Finally, the same experience was shown during a public art event to a few hundred participants. Many participants actively engaged with the game, especially family with kids. A short highlight video can be found at https://youtu.be/w6hnT7ZjdKg. We observed that kids found it easier than their parents to grasp the IFTT-PIN concept and adapt to the self-calibration paradigm. This is a testament to the incredible learning ability of young people, and a credible sign that users could become familiar with deciding what meaning they want to assign to buttons on an interface instead of trying to learn a pre-defined button-to-meaning mapping.

A more comprehensive study on usability and security with both quantitative and qualitative analysis is desirable to understand IFTT-PIN further. An interesting direction is to study if and how users can adapt to a self-calibrating interface and the role reversal it entails (i.e., deciding what meaning they want to assign to buttons on an interface instead of trying to learn a pre-defined button-to-meaning mapping), and to analyze the resulting variability of users' choices of action-to-meaning mapping. The vault setup is fully functional both in physical and digital form and would happily be made available to interested researchers upon request.

## 7 RELATED WORK

We review PIN-Entry methods resilient against shoulder surfing as well as interface similar to IFTT-PIN used in the field of brain-computer interface. We also review self-calibration work in both brain-computer interface and human-robot interaction and briefly refer to the field of experimental semiotics and language games which appeal to similar conceptual frameworks. This section is aimed to be a self-contained review, thus a few aspects covered in other sections might be repeated.

### 7.1 Human-Computer Interaction

#### *7.1.1 Shoulder surfing*

Shoulder surfing attacks happen when one observes other people's information without their explicit consent. PIN-entering is the stereotypical scenario considered and a large body of work focuses on PIN-entering methods that reduce the risk of shoulder surfing attacks. Common strategies include overwhelming the observer with visual decoys [3]–[5], indirect and/or clustered input [1], [5]–[7], and hidden secondary communication channels [8]–[10]. IFTT-PIN proposes a novel protection mechanism by removing the pre-established mapping between user action and their meaning. The user rather establishes this mapping on the fly and every time they use the interface, it makes it harder for malicious observers to interpret the user's actions.

Concretely, we implemented self-calibration principle on top of the interface used in the pioneering work of Roth et al. [1]. Their method works by splitting digits in two distinct sets, each set of digits visually identified by a color code (white and black), and then asking the user to click on the button of the color that is associated to the digit they want to enter. By repeating this process and shuffling the digit sets, the interface can identify the exact digit the user has in mind. IFTT-PIN adds self-calibration capabilities to the input button, meaning their color is not pre-defined but chosen by the user at interaction time, in their mind only and not communicated to the interface. This way, an outside observer will not know which color is associated to each button the user is pressing, making it much harder to infer the PIN. Usability and safety of this approach are discussed in section 6.

#### 7.1.2 Motion Matching

Under the motion matching paradigm, users select targets by mimicking their motion [11]. In most motion matching scenarios [12]–[14] there is no back-and-forth interaction between the user and the machine which we rely on for IFTT-PIN. However, pioneering work from Williamson et al. [15] consider an interactive motion matching scenario that shares many conceptual similarities with our work. First, their task is interactive as the user needs to actively control, rather than match, the motion of its chosen target out of many simultaneously visible targets. Second, their algorithms explicitly make use of interpretation hypotheses and they introduce variance as a consistency metric. Finally, by attributing a digit to each virtual agent, their experimental setup could readily be applied to PIN-entering and would be robust to shoulder surfing attacks.

One key difference with our work is that the action-to-meaning mapping is pre-defined in all motion-matching work known to us. For example, in [15], the mouse movements of the user directly influence the agent's motion in a predictable way. Applying self-calibration to their setup means we could, in theory, allow users to invent different mappings, e.g., moving in the same direction as their target rather than in the opposite direction. This would however add problems in the visual feedback loop to update the targets' positions in response to user's actions using a yet unknown action-to-motion relationship. This is a fascinating challenge for future research.

#### 7.1.3 Adaptive interfaces

An adaptive interface is an interface that can be changed to fit better to the end user needs. This can be done by changing the interface layout, adding buttons for specific functionality, but also by training an action-to-meaning mapping tailored to each user [16]. In most instances, the user can directly customize the interface via explicit protocols or hard-coded heuristics are used based on usage statistics, e.g., when reorganizing menu items depending on their frequency of usage. In some cases, the communication channel is fully learned from scratch prior to interaction. For example, in the Glove-Talk project [17], a mapping between user gestures and a speech synthesizer is learned by training a neural network based on examples collected from the user. However, this required the acquisition of a labeled dataset of user's actions associated to their meaning which can be impractical for the user and could require expert intervention.

IFTT-PIN proposes a new approach for developing adaptive interfaces where the user action-to-meaning mapping can be learned on the fly while the user is interacting with the device and without knowing in advance what they are trying to achieve. This process works well when the task has clear constraints, such as entering a PIN, but more research is required to scale to more complex tasks. Indeed, it is yet unclear how to scale self-calibration to a Glove-Talk-like scenario [17] with a large vocabulary of actions and free formed speech.

### 7.2 Brain-Computer Interfaces

Brain-computer interfaces (BCI) are relevant here because the control signals are not directly visible or intuitively interpretable to our senses. In addition, the bandwidth is very limited and only a few signals can be reliably generated by a user and decoded by a machine. These constraints led to creative interaction paradigms, such as the P300 speller, and, to our knowledge, drove the BCI community to develop the self-calibration paradigm.

#### 7.2.1 ERP and the P300 speller

Event-related potentials (ERPs) are stereotyped brain responses that are triggered by specific sensory-motor or cognitive events and can be measured via electroencephalography (EEG) [18]. The P300 event is a popular ERP because it is linked to active evaluation or categorization processes, induces a fast response (about 300ms after the event), and is reliably and reproducibly detectable [19], [20]. Thanks to these properties, researchers managed to harness P300 events to help human control artefacts.

The P300 speller was invented in the late 1980s [21] and allows patients to spell a text by iteratively choosing characters on a grid. The user looks at the character they want to type while characters are made to flash per row and per columns. Each time the chosen character is flashing, a specific P300 ERP response is triggered in the user's brain which can be detected using EEG. This technic has since evolved in many directions [20], but, fundamentally and of interest to us here, the P300 speller follows the exact same principle and protocol than the work of Roth et al. [1] and our work with IFTT-PIN.

Indeed, for the P300 speller, characters are split in two sets (flashing or not-flashing) and the user identifies the set associated to the chosen character by "generating" a P300 event (flashing or not-flashing). In Roth et al. [1], digits are split in two sets (white or black) and the user identifies the set associated to the chosen digit by pressing the corresponding button (white or black). In IFTT-PIN, we, too, use the same principle using yellow and grey instead of black and white.

#### *7.2.2 Self-calibration*

The P300 speller paradigm assumes access to a readily available decoder of brain signal into flashing/not-flashing events. This decoder needs to be tailored to each user during a calibration phase due to the variable and changing nature of the brain signals. During calibration, users are asked to spell specific letters to gather a labelled dataset on which a machine learning classifier can be trained. But this calibration process is tedious and requires the intervention of technical experts which adds costs and hinders the scalability of P300 systems.

The BCI community thus explored means to remove this calibration step altogether which, to our knowledge, led to the invention of the self-calibration paradigm. The main approach is to exploit redundancy or task constraints to narrow the possible set of decoders and use specific rules to prune this set and identify the true final decoder. For example, [22]–[24] exploit asymmetric repetition count to sort between hypotheses (one class has more members than the other by design), [25], [26] exploit the assumption that subjects intend to reach the target in straight line, [27], [28] adapt continuously a decoder to each individual starting from a generic weak decoder and [29], [30] directly rely on the subject's internal consistency by measuring similarity between signals of same meaning as they would be labelled for each hypothetical intent.

Further work has proposed an active planning algorithm to reduce the number of iterations required to identify the user intent under a self-calibration paradigm [2]. This is particularly challenging as the problem is entangled and both the user intent and their action-to-meaning mapping are unknown. Active planning is also implemented in IFTT-PIN and is the process of actively choosing which color pattern to apply on the digit with respect to the interaction history and in order to identify the user's digit faster.

The P300 speller and IFTT-PIN use the same mechanism of interaction, and the self-calibration paradigm was a direct source of inspiration for IFTT-PIN. Our contribution lies in adapting the algorithm presented in [2] from brain signals to discrete button events, and in adapting their active planning method to the discrete domain. Finally, we put considerable efforts in designing an approachable interface and made the entire system available as an interactive demo for the community to try and share.

### 7.3 Human-Robot Interaction

Interactive learning is the process of learning a task by practical interaction with a user [31]. In robotics, this process is useful to solve complex tasks that require multi-step planning and are hard to describe and solve via handcrafted rules. A usual assumption is that the human provides instructions following a known protocol and using teaching signals the robot already knows how to interpret. In [32], Lopes at al. proposed a method to allow the robot to extend its vocabulary over time starting from a set of known signs. In [33], Najar et al. demonstrates the use of unlabelled human instructions to interactively shape a robot's behavior in a reinforcement learning scenario. No starting vocabulary is required but the robot has access to a reward signal informing it when the final goal has been achieved. Because there are many ways to achieve the same task, the user's unlabelled instructions help to shape the robot towards a preferred policy, as well as to accelerate learning. In [34], Grizou et al. removed the need of a known starting vocabulary or access to a reward function, effectively creating a self-calibrating system on a pick-and-place robotic task with a human providing instructions using spoken words.

### 7.4 Language Acquisition

Looking into the study of language acquisition is also relevant for this work because it also features a chicken-and-egg problem. How can we acquire language when we do not yet know how to understand others? Below, we briefly refer related work in experimental semiotics and computation modelling of language acquisition.

#### *7.4.1 Experimental Semiotics*

Experimental semiotics [35] studies the emergence and evolution of novel communication systems by conducting controlled studies with human adults. We point interested readers to three works [36]–[38] that consider asymmetric communication. In such experiments, only one participant can send communicative signals to an observer and the observer goal is to understand the intention of that person. But both the exact task to be solved and the communicative signals are initially unknown. To make the parallel clear, in IFTT-PIN experiments, one participant, the user, sends communicative signals to the machine (button presses), and the machine should infer the user's PIN. But both the action-to-meaning mapping and the PIN of the user are initially unknown.

#### *7.4.2 Language Games*

Researchers studying the origin and evolution of language often rely on computational simulation of populations of artificial agents [39]. Typically, agents randomly interact with each other's and play language games requiring some form of communication, but no pre-agreed language exists between agents. Running these simulations helps study how a language is created and propagates within a population. Chicken-and-egg situations are often solved via explicit disambiguation protocols inherent to the game played between agents (e.g. guessing game [40], categorization game [41]). The work of Cederborg et al. [42] is most relevant for us because, like IFTT-PIN, they rely on a combination of interpretation hypotheses and consistency assumptions to solve their language acquisition dilemma.

## 8 LIMITATIONS AND FUTURE WORK

In its current implementation, IFTT-PIN assumes users are perfect and do not make any mistakes. Human errors could be accounted for in two ways: (1) adding an explicit interface to undo the last action, and (2) modeling an error rate in the self-calibration algorithm, similar to the implementation in [2]. This would allow users to make a certain amount of error but would also increase the number of clicks required to enter a digit. Similarly, the current interface lacks feedback mechanisms to the user with respect to the advancement of the digit identification process. While we show the color of the buttons and each digit once they are successfully identified, the user has no way of knowing how far along they are in the identification process of each digit. Showing a progress bar is a potential solution but might not be straightforward to implement as the number of remaining iterations is not knowable with certainty lacking a known action-to-meaning mapping.

Regarding shoulder surfing attacks, although IFTT-PIN is more robust than a traditional PIN pad and adds an additional layer of security on top of the work of Roth et al. [1] by removing the visible and pre-defined color assignment on each button, a number of routes for improvement could be investigated. As already suggested in Roth et al. [1], we could introduce a probabilistic trap door approach were the user does not need to enter the exact PIN to open the vault, but simply to reduce the set of possible PINs to a small number that includes the correct PIN. That way, an attacker would only be able to recover the set, not knowing which specific PIN in that set is correct, thus not being certain which one to choose to replicate. Another approach would be to introduce a hidden communication channel that only the user can access. For example, we could equip the user with special glasses that reveal to their users the color assigned to each digit. Outside observers or cameras would not see the color on the digit, making it impossible to reverse-engineer what PIN the user is entering.

In terms of interaction modalities, the current version of IFTT-PIN only considers discrete button presses events which makes it simple to measure consistency as using "the same buttons for the same meanings". It would be interesting to build a version of IFTT-PIN accepting continuous input signals, for example allowing the user to use different gestures to mean yellow and grey. The main challenge is that two gestures for the same meaning will never be performed twice in exactly the same way. Detecting inconsistencies would thus be significantly harder than when using buttons which would require establishing a softer consistency metric of "similar gestures used for the same meanings". This can be approached using the algorithms developed in the BCI community as discussed in section 7.2, and will be part of our future work.

But the main challenges ahead lie with scaling IFTT-PIN and the self-calibration paradigm to more complex problems. With the PIN-entering task, we are working on a very limited set of possible intent, meaning, and action which allow for the interpretation hypothesis process to be computationally tractable. Most of the literature has not scaled beyond this point and considering continuous tasks, a larger set of meanings and actions is a difficult challenge that will require novel approaches to be invented.

Finally, beyond the technical challenges, it remains unclear where self-calibrating interfaces could be genuinely beneficial in our everyday world. The shoulder surfing context is a niche one, it is not a common problem for PIN entry tasks and does not usually have adverse consequences [43]. The BCI context is the most convincing so far as EEG signals cannot be easily observed and are changing both in time and from person to person which means repeated calibration is a necessity. However, in some cases, repeated expert assistance for calibration might be too inconvenient or expensive and hinder user's access to these technologies. Furthermore, under severe handicap, a P300 speller paradigm, while slow, remains or is the best possible communication paradigm available. Which modalities and what applications could benefit from having self-calibrating interfaces remains a widely open question and is the one we wish to ask via IFTT-PIN.

## 9 CONCLUSION

We presented IFTT-PIN, an online interactive PIN-entry interface conceived as a vehicle to introduce the self-calibration paradigm. IFTT-PIN allows users to enter the PIN of their choice via an elimination process by indicating the color assigned to their digit (yellow or grey). To express their choice, users should click on a button whose color is the same as their digit. But buttons do not have any color assigned to them at the start of the interaction. Users are free to define the color of each button in their mind and use them as such without informing the interface. After a few iterations, the interface can infer both the digit the user had in mind and the color of each button used. IFTT-PIN can thus effectively calibrate itself to each user at interaction time, demonstrating a new interactive experience.

For this reason, we propose IFTT-PIN as a novel approach to protect against shoulder surfing attacks. Indeed, because the color of each button is not pre-defined and only pre-exists in the user's mind, it is much harder for an attacker to make sense of the actions of the user in real time. We further introduced a vault opening challenge to assess the usability and security of IFTT-PIN. Preliminary informal observations indicate that the interface is both usable (participants managed to open the vault), secure (it took them hours to crack the code with access to a recorded video of PIN-entry) and is an engaging challenge for educational and experimental purposes.

Future work will address the issue of scalability to more complex input modalities (e.g., gestures, voice), varied interaction protocol (e.g., guidance instead of feedback) and more complex tasks (e.g., continuous set of goals). We are also interested in studying the diversity of users' choice of action-to-meaning mapping under the self-calibration paradigm, and whether it can effectively lead to more varied and personalized interaction schemes.

Lastly, we would like to encourage researchers to investigate what other interfaces and applications could benefit from the self-calibration paradigm - especially towards more subjective interactive devices where users can express their intention in an intuitive, fluid and personalized way without resorting to explicit calibration procedures. We hope this work will inspire the community to invent them.

## ACKNOWLEDGMENTS

We thank the CRI for the fellowship that allowed us to develop this demo. Many thanks to Edwin Paquiot and Joanna May Lee for their design advice, as well as to students and staff at CRI for testing this demo extensively. This work was submitted as UIST 2021 (https://uist.acm.org/uist2021/) and not accepted due to the lack of user testing. We believe including such tests would anchor the debate around the shoulder-surfing applicability and focus readers on comparing IFTT-PIN to alternatives based on various metrics. Our goal with this work is solely to deliver an introductory experience to the self-calibrating paradigm, shoulder-surfing is secondary and used for framing around a familiar PIN-entry task.

## REFERENCES

[1] V. Roth, K. Richter, and R. Freidinger, "A PIN-entry method resilient against shoulder surfing," in *Proceedings of the 11th ACM conference on Computer and communications security*, New York, NY, USA, Oct. 2004, pp. 236–245, doi: 10.1145/1030083.1030116.

[2] J. Grizou, I. Iturrate, L. Montesano, P.-Y. Oudeyer, and M. Lopes, "Interactive learning from unlabeled instructions," in *Proceedings of the Thirtieth Conference on Uncertainty in Artificial Intelligence*, Arlington, Virginia, USA, Jul. 2014, pp. 290–299, Accessed: Feb. 02, 2021. [Online].

[3] J. Gugenheimer, A. De Luca, H. Hess, S. Karg, D. Wolf, and E. Rukzio, "ColorSnakes: Using Colored Decoys to Secure Authentication in Sensitive Contexts," in *Proceedings of the 17th International Conference on Human-*

*Computer Interaction with Mobile Devices and Services*, New York, NY, USA, Aug. 2015, pp. 274–283, doi: 10.1145/2785830.2785834.

[4] D. S. Tan, P. Keyani, and M. Czerwinski, “Spy-resistant keyboard: more secure password entry on public touch screen displays,” in *Proceedings of the 17th Australia conference on Computer-Human Interaction: Citizens Online: Considerations for Today and the Future*, Narrabundah, AUS, Nov. 2005, pp. 1–10, Accessed: Mar. 08, 2021. [Online].

[5] A. De Luca, K. Hertzschuch, and H. Hussmann, “ColorPIN: securing PIN entry through indirect input,” in *Proceedings of the SIGCHI Conference on Human Factors in Computing Systems*, New York, NY, USA, Apr. 2010, pp. 1103–1106, doi: 10.1145/1753326.1753490.

[6] E. von Zezschwitz, A. De Luca, B. Brunkow, and H. Hussmann, “SwiPIN: Fast and Secure PIN-Entry on Smartphones,” in *Proceedings of the 33rd Annual ACM Conference on Human Factors in Computing Systems*, New York, NY, USA, Apr. 2015, pp. 1403–1406, doi: 10.1145/2702123.2702212.

[7] W. A. J. van Eekelen, J. van den Elst, and V.-J. Khan, “Picassopass: a password scheme using a dynamically layered combination of graphical elements,” in *CHI ’13 Extended Abstracts on Human Factors in Computing Systems*, New York, NY, USA, Apr. 2013, pp. 1857–1862, doi: 10.1145/2468356.2468689.

[8] A. Bianchi, I. Oakley, V. Kostakos, and D. S. Kwon, “The phone lock: audio and haptic shoulder-surfing resistant PIN entry methods for mobile devices,” in *Proceedings of the fifth international conference on Tangible, embedded, and embodied interaction*, New York, NY, USA, Jan. 2010, pp. 197–200, doi: 10.1145/1935701.1935740.

[9] A. De Luca *et al.*, “Now you see me, now you don’t: protecting smartphone authentication from shoulder surfers,” in *Proceedings of the SIGCHI Conference on Human Factors in Computing Systems*, New York, NY, USA, Apr. 2014, pp. 2937–2946, doi: 10.1145/2556288.2557097.

[10] M. Khamis, F. Alt, M. Hassib, E. von Zezschwitz, R. Hasholzner, and A. Bulling, “GazeTouchPass: Multimodal Authentication Using Gaze and Touch on Mobile Devices,” in *Proceedings of the 2016 CHI Conference Extended Abstracts on Human Factors in Computing Systems*, New York, NY, USA, May 2016, pp. 2156–2164, doi: 10.1145/2851581.2892314.

[11] J.-D. Fekete, N. Elmqvist, and Y. Guiard, “Motion-pointing: target selection using elliptical motions,” in *Proceedings of the SIGCHI Conference on Human Factors in Computing Systems*, New York, NY, USA, Apr. 2009, pp. 289–298, doi: 10.1145/1518701.1518748.

[12] D. Verweij, A. Esteves, V.-J. Khan, and S. Bakker, “WaveTrace: Motion Matching Input using Wrist-Worn Motion Sensors,” in *Proceedings of the 2017 CHI Conference Extended Abstracts on Human Factors in Computing Systems*, New York, NY, USA, May 2017, pp. 2180–2186, doi: 10.1145/3027063.3053161.

[13] M. Vidal, A. Bulling, and H. Gellersen, “Pursuits: spontaneous interaction with displays based on smooth pursuit eye movement and moving targets,” in *Proceedings of the 2013 ACM international joint conference on Pervasive and ubiquitous computing*, New York, NY, USA, Sep. 2013, pp. 439–448, doi: 10.1145/2493432.2493477.

[14] M. Carter, E. Velloso, J. Downs, A. Sellen, K. O’Hara, and F. Vetere, “PathSync: Multi-User Gestural Interaction with Touchless Rhythmic Path Mimicry,” in *Proceedings of the 2016 CHI Conference on Human Factors in Computing Systems*, New York, NY, USA, May 2016, pp. 3415–3427, doi: 10.1145/2858036.2858284.

[15] J. Williamson and R. Murray-Smith, “Pointing without a pointer,” in *CHI ’04 Extended Abstracts on Human Factors in Computing Systems*, New York, NY, USA, Apr. 2004, pp. 1407–1410, doi: 10.1145/985921.986076.

[16] D. Browne, *Adaptive User Interfaces*. Elsevier, 2016.

[17] S. S. Fels and G. E. Hinton, “Glove-talk II - a neural-network interface which maps gestures to parallel formant speech synthesizer controls,” *IEEE Transactions on Neural Networks*, vol. 8, no. 5, pp. 977–984, Sep. 1997, doi: 10.1109/72.623199.

[18] S. J. Luck, *An Introduction to the Event-Related Potential Technique, second edition*. MIT Press, 2014.

[19] P. Tw, “The P300 wave of the human event-related potential.,” *J Clin Neurophysiol*, vol. 9, no. 4, pp. 456–479, Oct. 1992, doi: 10.1097/00004691-199210000-00002.

[20] A. Kawala-Sterniuk *et al.*, “Summary of over Fifty Years with Brain-Computer Interfaces—A Review,” *Brain Sciences*, vol. 11, no. 1, Art. no. 1, Jan. 2021, doi: 10.3390/brainsci11010043.

[21] L. A. Farwell and E. Donchin, “Talking off the top of your head: toward a mental prosthesis utilizing event-related brain potentials,” *Electroencephalography and Clinical Neurophysiology*, vol. 70, no. 6, pp. 510–523, Dec. 1988, doi: 10.1016/0013-4694(88)90149-6.

[22] D. Hübner, T. Verhoeven, K. Schmid, K.-R. Müller, M. Tangermann, and P.-J. Kindermans, "Learning from label proportions in brain-computer interfaces: Online unsupervised learning with guarantees," *PLOS ONE*, vol. 12, no. 4, p. e0175856, Apr. 2017, doi: 10.1371/journal.pone.0175856.
[23] P.-J. Kindermans, D. Verstraeten, and B. Schrauwen, "A Bayesian Model for Exploiting Application Constraints to Enable Unsupervised Training of a P300-based BCI," *PLOS ONE*, vol. 7, no. 4, p. e33758, Apr. 2012, doi: 10.1371/journal.pone.0033758.
[24] P.-J. Kindermans, M. Schreuder, B. Schrauwen, K.-R. Müller, and M. Tangermann, "True Zero-Training Brain-Computer Interfacing – An Online Study," *PLOS ONE*, vol. 9, no. 7, p. e102504, Jul. 2014, doi: 10.1371/journal.pone.0102504.
[25] A. L. Orsborn, S. Dangi, H. G. Moorman, and J. M. Carmena, "Closed-Loop Decoder Adaptation on Intermediate Time-Scales Facilitates Rapid BMI Performance Improvements Independent of Decoder Initialization Conditions," *IEEE Transactions on Neural Systems and Rehabilitation Engineering*, vol. 20, no. 4, pp. 468–477, Jul. 2012, doi: 10.1109/TNSRE.2012.2185066.
[26] A. L. Orsborn, H. G. Moorman, S. A. Overduin, M. M. Shanechi, D. F. Dimitrov, and J. M. Carmena, "Closed-Loop Decoder Adaptation Shapes Neural Plasticity for Skillful Neuroprosthetic Control," *Neuron*, vol. 82, no. 6, pp. 1380–1393, Jun. 2014, doi: 10.1016/j.neuron.2014.04.048.
[27] A. Barachant and M. Congedo, "A Plug&Play P300 BCI Using Information Geometry," *arXiv:1409.0107 [cs, stat]*, Aug. 2014, Accessed: Mar. 11, 2021. [Online]. Available: http://arxiv.org/abs/1409.0107.
[28] C. Vidaurre, M. Kawanabe, P. von Bünau, B. Blankertz, and K. R. Müller, "Toward Unsupervised Adaptation of LDA for Brain–Computer Interfaces," *IEEE Transactions on Biomedical Engineering*, vol. 58, no. 3, pp. 587–597, Mar. 2011, doi: 10.1109/TBME.2010.2093133.
[29] J. Grizou, I. Iturrate, L. Montesano, P.-Y. Oudeyer, and M. Lopes, "Calibration-Free BCI Based Control," *AAAI*, vol. 28, no. 1, Art. no. 1, Jun. 2014, Accessed: Feb. 02, 2021. [Online]. Available: https://ojs.aaai.org/index.php/AAAI/article/view/8923.
[30] I. Iturrate, J. Grizou, J. Omedes, P.-Y. Oudeyer, M. Lopes, and L. Montesano, "Exploiting Task Constraints for Self-Calibrated Brain-Machine Interface Control Using Error-Related Potentials," *PLOS ONE*, vol. 10, no. 7, p. e0131491, Jul. 2015, doi: 10.1371/journal.pone.0131491.
[31] C. Breazeal *et al.*, "Tutelage and collaboration for humanoid robots," *Int. J. Human. Robot.*, vol. 01, no. 02, pp. 315–348, Jun. 2004, doi: 10.1142/S0219843604000150.
[32] M. Lopes, T. Cederbourg, and P. Oudeyer, "Simultaneous acquisition of task and feedback models," in *2011 IEEE International Conference on Development and Learning (ICDL)*, Aug. 2011, vol. 2, pp. 1–7, doi: 10.1109/DEVLRN.2011.6037359.
[33] A. Najar, O. Sigaud, and M. Chetouani, "Interactively shaping robot behaviour with unlabeled human instructions," *Auton Agent Multi-Agent Syst*, vol. 34, no. 2, p. 35, May 2020, doi: 10.1007/s10458-020-09459-6.
[34] J. Grizou, M. Lopes, and P. Oudeyer, "Robot learning simultaneously a task and how to interpret human instructions," in *2013 IEEE Third Joint International Conference on Development and Learning and Epigenetic Robotics (ICDL)*, Aug. 2013, pp. 1–8, doi: 10.1109/DevLrn.2013.6652523.
[35] B. Galantucci and S. Garrod, "Experimental Semiotics: A Review," *Front Hum Neurosci*, vol. 5, Feb. 2011, doi: 10.3389/fnhum.2011.00011.
[36] J. P. de Ruiter *et al.*, "Exploring the cognitive infrastructure of communication," *IS*, vol. 11, no. 1, pp. 51–77, Mar. 2010, doi: 10.1075/is.11.1.05rui.
[37] S. Griffiths *et al.*, "Bottom-up learning of feedback in a categorization task," in *2012 IEEE International Conference on Development and Learning and Epigenetic Robotics (ICDL)*, San Diego, CA, USA, Nov. 2012, pp. 1–6, doi: 10.1109/DevLrn.2012.6400864.
[38] A. Vollmer, J. Grizou, M. Lopes, K. Rohlfing, and P. Oudeyer, "Studying the co-construction of interaction protocols in collaborative tasks with humans," in *4th International Conference on Development and Learning and on Epigenetic Robotics*, Oct. 2014, pp. 208–215, doi: 10.1109/DEVLRN.2014.6982983.
[39] L. Steels, "The Synthetic Modeling of Language Origins," *EOC*, vol. 1, no. 1, pp. 1–34, Jan. 1997, doi: 10.1075/eoc.1.1.02ste.
[40] L. Steels, "Language games for autonomous robots," *IEEE Intelligent Systems*, vol. 16, no. 5, pp. 16–22, Sep. 2001, doi: 10.1109/MIS.2001.956077.

[41] V. Loreto, A. Baronchelli, A. Mukherjee, A. Puglisi, and F. Tria, “Statistical physics of language dynamics,” *J. Stat. Mech.*, vol. 2011, no. 04, p. P04006, Apr. 2011, doi: 10.1088/1742-5468/2011/04/P04006.
[42] T. Cederborg and P. Oudeyer, “Imitating operations on internal cognitive structures for language aquisition,” in *2011 11th IEEE-RAS International Conference on Humanoid Robots*, Oct. 2011, pp. 650–657, doi: 10.1109/Humanoids.2011.6100875.
[43] M. Eiband, M. Khamis, E. von Zezschwitz, H. Hussmann, and F. Alt, “Understanding Shoulder Surfing in the Wild: Stories from Users and Observers,” in *Proceedings of the 2017 CHI Conference on Human Factors in Computing Systems*, Denver Colorado USA, May 2017, pp. 4254–4265, doi: 10.1145/3025453.3025636.